\documentclass[thmsa,notitlepage,11pt]{article}
\usepackage{amssymb}
\usepackage{amsfonts}
\usepackage{sw20lart}

\input{tcilatex}

\begin{document}

\title{{\LARGE Less suppressed }$\mu \rightarrow e\gamma $ {\LARGE and} $%
\tau \rightarrow \mu \gamma ${\LARGE \ loop amplitudes and extra dimension
theories}}
\author{{\large Bo He}$^{\diamond }$\thanks{%
Present address: Department of Electric Engineering, Ohio State University,
Columbus, OH 43210}~{\large , T. P. Cheng}$^{\diamond }$\thanks{%
Corresponding author: tpcheng@umsl.edu, 1-314-516-5020 (phone), -6152 (fax) }%
{\large \ }and{\large \ Ling-Fong Li}$^{\ddagger }$ \\
%EndAName
$^{\diamond }${\small Department of Physics and Astronomy, University of
Missouri, St. Louis, MO 63121}\\
$^{\ddagger }${\small Department of Physics, Carnegie Mellon University,
Pittsburgh, PA 15213}}
\maketitle

\begin{abstract}
When $\mu e\gamma $ (or $\tau \mu \gamma $) loop involves a vector boson,
the amplitude is suppressed by more than two powers of heavy particle
masses. However we show that the scalar boson loop diagrams are much less
damped. Particularly, the loop amplitude in which the intermediate fermion
and scalar boson have comparable masses is as large as possible, as allowed
by the decoupling theorem. Such a situation is realized in the "universal
extra dimension theory", and can yield a large enough rate for $\mu e\gamma $
to be detectable in current experiments. Our investigation involves precise
calculation of the scalar boson loop's dependence on the masses of the
intermediate states.

\bigskip 


\end{abstract}

\newpage

\section{Introduction}

\paragraph{Probing the physics beyond the SM through the $\protect\mu e%
\protect\gamma $ loop effect}

The Standard Model with massless neutrinos automatically conserves lepton
flavors: the electron, muon and tau numbers. The ever stronger experimental
evidence for neutrino oscillation\cite{nu-oscillation} shows clearly that
lepton flavor is not conserved in nature. If we accommodate this feature
simply by an introduction of neutrino masses in the SM, other lepton flavor
violating processes such as $\mu \rightarrow e\gamma $ would still have so
small a rate (a branching ratio $\ll 10^{-40}$) that there is no hope for
their detection in the foreseeable future. This is the case whether we have
Dirac neutrino masses with their small values inserted by hand, or the
neutrinos are Majorana particles with the smallness of their masses coming
out of the seesaw mechanism. In the small Dirac mass scenario, the $\mu
e\gamma $ amplitude is suppressed by the neutrino mass difference $\delta
m_{\nu }^{2}$ over the vector boson mass $M_{W}^{2}$, namely, suppressed by
a leptonic GIM scheme\cite{mu-e-g-original}. In the seesaw scenario,
superheavy singlet neutrino states are present in the usual left-handed
flavor eigenstates. Their potentially significant contribution is
nevertheless muffled by the mass-suppressed mixing angles\cite{CL01}. Thus a
detection of the $\mu e\gamma $ process would signal the physics beyond the
SM, beyond any neutrino mixing mechanism.

We are particularly interested in the possibility of this $\mu e\gamma $
amplitude being less damped in theories that predict new particles around $%
TeV$ \cite{kitano}\cite{CL01}. One is curious whether the contribution by
such heavy particles to the $\mu e\gamma $ amplitude could be as large as
possible while remaining compatible with the decoupling theorem\cite%
{app-carr-thm}. For example, the suppressions mentioned in the last
paragraph both involve at least two powers of heavy particle mass in the
denominator. Are there situations in which the amplitude suppression is
linear? In this paper we report a precise calculation of the $\mu e\gamma $
amplitude's dependence on its intermediate particle masses for the scalar
boson loop diagrams. It shows that, when the intermediate fermion and boson
masses are comparable, such a potentially detectable $\mu e\gamma $ rate is
possible. While our result is applicable in any theory that allows this
lepton flavor nonconserving decay, in this paper we\ present it mainly in
the context of "large extra dimension theories" as their phenomenology has
lately been under active discussion.

\paragraph{Large extra dimension theories}

In the last few years there has been considerable interest in theories
having extra dimensions, which either have their compactification scales
being much larger than the Planck's length\cite{largeKKoriginal}, or have
strong curvature\cite{RS}. These theories can in principle generate the
observed gauge hierarchy, for example, by having large extra dimensional
volume. Such theoretical suggestion would also provide an added impetus for
the ongoing experimental effort to test Newton's gravity theory at the
millimeter scale.

The hallmark of extra dimensions is the existence of Kaluza-Klein (KK)
states: Particles that can propagate in the compactified extra dimension
have a tower of states with identical quantum numbers but ever increasing
masses. For the simplest case of a scalar particle in a five dimensional
spacetime, with the extra dimension compactified into a circle (radius $R$),
we can expand the field $\phi \left( x^{\mu },x^{5}\right) $ in harmonics 
\begin{equation}
\phi \left( x^{\mu },x^{5}\right) =\sum_{n}\phi _{n}\left( x^{\mu }\right)
e^{ip_{5}x^{5}}.  \label{harmonics}
\end{equation}%
Because the extra dimension is a circle, the positions $x^{5}$ and $%
x^{5}+2\pi R$ are identified, with $\phi \left( x^{\mu },x^{5}+2\pi R\right)
=\phi \left( x^{\mu },x^{5}\right) $ leading to the quantization of the
extra dimension momentum: $p_{5}=n/R,$ where\ $n=0,1,2,...$ is the KK
number. This implies a mass spectrum of $M^{2}\left( n\right) =M^{2}\left(
0\right) +n^{2}/R^{2}$, where the zero mode bare mass $M\left( 0\right) $ is
expected to be much smaller than the KK excitation energy of $1/R.$ Thus
there would be a tower of KK states associated with any particle that can
propagate in an extra dimension.

\paragraph{Brane \emph{vs} bulk particles: the universal extra dimension
theory}

Different extra dimension theories often have different particle assignments
vis-a-vis whether they can propagate in the full higher dimensional space or
not. Those do are "bulk particles", and have associated KK states, while the
"brane particles" are those confined (on the brane) to the four dimensional
spacetime. The original large dimension theory has all the SM particles
stuck on the brane, and only graviton is a bulk field\cite{largeKKoriginal}.
Prior investigation\cite{old-extraD} and later variations include
suggestions in which the neutrinos\cite{bulk-nu}, the scalar bosons, or the
vector gauge bosons\cite{bulk-gauge}, etc. propagate in the extra dimensions
as well. Furthermore, with the presence of brane, the translational
invariance in the extra dimensional space is broken, and the corresponding
extra dimension momentum (the KK number) is not conserved.

Among the different modifications of the original large extra dimension
model,\ the most appealing suggestion, to our thinking, has been that by
Appelquist, Cheng and Dobrescu\cite{univKK} who proposed that \emph{all}
standard model particles, as well as graviton, can propagate in the extra
dimensions, thus all particles have KK states. These are "universal extra
dimensions". Because brane's presence is no longer required, translational
invariance, and KK number conservation, are restored. There are no vertices
involving only one non-zero KK mode. This\ is the key feature that allows
such a large extra dimensional theory to pass all the phenomenological
tests. Appelquist \emph{et al}. analyzed the current electroweak data,
computed some parameters and concluded that the compactification size $R$
could be $1/300GeV$ for one extra dimension and $1/400\rightarrow 1/800GeV$
for two extra dimensions. These predictions are in the range of current or
near-future experiments. Other discussions about the experiment signatures
of the universal extra dimensions can also be found in the literature\cite%
{univKK-others}

\paragraph{KK particles in the $\protect\mu e\protect\gamma $ loop}

We are interested in estimating the $\mu e\gamma $ rate as induced by the
loop diagram in which the intermediate virtual particles (one fermion and
one boson) may be KK states. We shall broadly distinguish two categories of
models: In one category, theories do not have KK number conservation, and in
such models it's possible that only one of the virtual particles is a heavy
KK state. Here the general situation corresponds to mass limits when either
the fermion mass is much larger than the boson mass, or the other way
around. In the second category, we consider the KK number conserving
universal extra dimension theory. Here both the fermion and boson must have
the same KK number and their masses are comparable. This is so, because
their mass square difference is the same as that between their zero modes,
which is expected to be much smaller than the KK excitation. In order to
consider the comparable mass case, we will have to compute the one loop
amplitude exactly in its dependence of the intermediate particle mass ratio.

The gauge invariant decay amplitude for $\mu \left( p\right) \rightarrow
e\left( p-q\right) +\gamma \left( q,\epsilon \right) $ must have the form 
\begin{equation}
T\left( \mu e\gamma \right) =\frac{ie}{16\pi }\varepsilon _{\lambda }^{\ast
}\left( q\right) \bar{u}_{e}\left( p-q\right) \sigma ^{\lambda \rho }q_{\rho
}\left[ A_{+}\left( 1+\gamma _{5}\right) +A_{-}\left( 1-\gamma _{5}\right) %
\right] u_{\mu }\left( p\right) .  \label{amplitude}
\end{equation}%
This corresponds to a dimension-five Lagrangian density term $\bar{\psi}%
_{e}\sigma ^{\lambda \rho }\psi _{\mu }F_{\lambda \rho }.$ The invariant
amplitudes $A_{\pm }$ are induced by finite and calculable loop diagrams and
proportional to an inverse mass power.

\section{Vector loop amplitude}

In the SM with one doublet of Higgs bosons and small Dirac neutrino masses,
there is only one type of loop diagrams (Fig 1) for the $\mu e\gamma $
decay. They have a charged intermediate boson in the loop: $\mu
^{-}\rightarrow \left( \nu _{i}W_{\gamma }^{-}\right) \rightarrow e^{-}$
where the photon is emitted by the charged $W$ boson in the loop (as denoted
by the subscript $\gamma $).

The required exact mass calculation has been performed\cite{CL-ps00}\cite%
{kitano}\cite{CL01} giving, in the $m_{e}=0$ approximation, the amplitudes
of $A_{-}^{W}=0$ and 
\begin{equation}
A_{+}^{W}=\frac{g^{2}m_{\mu }}{8\pi M_{W}^{2}}\sum_{i=1}^{3}\mathbb{U}_{\mu
i}^{\ast }\mathbb{U}_{ei}F\left( \frac{m_{i}^{2}}{M_{W}^{2}}\right) ,
\label{W-amplitude}
\end{equation}%
where the function\footnote{%
The sign in front of the log term was incorrectly written down in \cite{CL01}%
.} 
\begin{equation}
F\left( z\right) =\frac{1}{6\left( 1-z\right) ^{4}}\left(
10-43z+78z^{2}-49z^{3}+18z^{3}\ln z+4z^{4}\right) .  \label{F-z}
\end{equation}%
has limits%
\begin{eqnarray}
F(z &\rightarrow &\infty )\simeq \frac{2}{3}+3\frac{\ln z}{z}
\label{limit-a} \\
&&  \nonumber \\
F(z &\rightarrow &0)\simeq \frac{5}{3}-\frac{1}{2}z  \label{limit-b} \\
&&  \nonumber \\
F(z &\rightarrow &1)\simeq \frac{17}{12}+\frac{3}{20}\left( 1-z\right)
\label{limit-c}
\end{eqnarray}%
The resultant amplitudes, after using the unitarity condition of the mixing
matrices $\sum_{i=1}^{3}\mathbb{U}_{ei}^{\ast }\mathbb{U}_{\mu i}=0,$ are
summarized in Table 1.%
\begin{eqnarray*}
&&\text{ \ \ \ \ \ \ \ \ \ \ \ Table 1. \ The vector loop amplitudes }%
A_{+}^{W}~ \\
&&%
\begin{tabular}{|c|c|}
\hline
limits & $A_{+}^{W}$ \\ \hline
$m_{i}\gg M_{W}$ & $\ \ \ \ \ \ \ \frac{3g^{2}m_{\mu }}{8\pi }%
\sum_{i=1}^{3}\left( \frac{1}{m_{i}^{2}}\ln \frac{m_{i}^{2}}{M_{W}^{2}}%
\right) \mathbb{U}_{\mu i}^{\ast }\mathbb{U}_{ei}\ \ \ \ \ \ $ \\ 
&  \\ 
$m_{i}\ll M_{W}$ & $\ \ \ \ \ \ \ \ -\frac{g^{2}m_{\mu }}{16\pi M_{W}^{4}}%
\sum_{i=1}^{3}m_{i}^{2}\mathbb{U}_{\mu i}^{\ast }\mathbb{U}_{ei}\ \ \ \ \ \ $
\\ 
&  \\ 
$m_{i}\simeq M_{W}$ & $\ \ \ \ \ \ \ \ \frac{3g^{2}m_{\mu }}{160\pi M_{W}^{2}%
}\sum_{i=1}^{3}\frac{M_{W}^{2}-m_{i}^{2}}{M_{W}^{2}}\mathbb{U}_{\mu i}^{\ast
}\mathbb{U}_{ei}\ \ \ \ \ \ $ \\ \hline
\end{tabular}%
\end{eqnarray*}

The limit $m_{i}\gg M_{W}$ is relevant to models in which the neutrino is a
bulk field while the vector boson\ $W$\ is not. The amplitude $A_{+}^{W}$ is
suppressed by the heavy mass as $\left( m_{\mu }/m_{i}^{2}\right) \ln m_{i}$%
. The $m_{i}\ll M_{W}$ case includes the specific situation\ of massless
neutrinos $m_{i}=0,$ which leads to a vanishing amplitude, as lepton flavor
must be conserved in the massless neutrino limit. For neutrinos with small
(zero mode) masses $\left( m_{i}\right) _{0}\ll \left( M_{W}\right) _{0}$,
as well as the case when only $W$ has KK states, the amplitude is
proportional to neutrino mass-squared $m_{\mu }m_{i}^{2}/M_{W}^{4}.$ This
results in a branching ratio so small that the decay cannot be detected
experimentally in the foreseeable future. One might think that the situation
of $m_{i}\simeq M_{W}$ could offer a better chance of having a less
suppressed amplitude. But this turns out not to be so. Since, for a given KK
number, the mass-square-difference is given by that of the zero modes $%
\left( M_{W}^{2}-m_{i}^{2}\right) _{n\neq 0}=\left(
M_{W}^{2}-m_{i}^{2}\right) _{0}$, the amplitude is again suppressed by $%
m_{\mu }\left( m_{i}^{2}\right) _{0}/\left( M_{W}^{4}\right) _{n}$ leading
to an undetectably small branching ratio.

\section{Scalar loop amplitude}

Although the minimal SM needs only one doublet of Higgs particles, in most
extensions more Higgs doublets are introduced. For example, in supersymmetry
two Higgs doublets with opposite hypercharges are required. Several versions
of compactification of superstring theory leads to E$_{6}$ grand unified
theories where each generation of leptons and quarks has a pair of
oppositely hypercharged Higgs scalar boson. Thus there is strong motivation
to consider theories with multiples of scalar bosons. Here we are interested
how such scalars, and their possible KK excitations, can contribute to the $%
\mu e\gamma $ loop amplitude\cite{BJweinberg}. Just as in the vector loop
case, there is the need to obtain the exact intermediate mass dependence. We
have performed this task and obtained the following result.

In Fig 2(a) we have intermediate states of a charged scalar boson and a
neutrino. Denote the Yukawa couplings of the scalar boson $\phi $ to leptons 
$\emph{l}_{i}$ and $l_{j}$ by $y_{li}^{\pm }$, 
\begin{equation}
\Gamma \left( \phi l_{i}l_{j}\right) =\bar{l}_{i}\left[ y_{ij}^{+}\left(
1+\gamma _{5}\right) +y_{ij}^{-}\left( 1-\gamma _{5}\right) \right]
l_{j}\phi +h.c.
\end{equation}%
we find the amplitudes to be%
\begin{eqnarray}
A_{+}^{\phi }(a) &=&-\sum_{i}\frac{m_{\mu }}{\pi M_{\phi }^{2}}%
y_{ei}^{+}y_{\mu i}^{-}G(r)-\sum_{i}\frac{m_{i}}{\pi M_{\phi }^{2}}%
y_{ei}^{+}y_{\mu i}^{+}I(r)  \label{A+a} \\
&&  \nonumber \\
A_{-}^{\phi }(a) &=&-\sum_{i}\frac{m_{\mu }}{\pi M_{\phi }^{2}}%
y_{ei}^{-}y_{\mu i}^{+}G(r)-\sum_{i}\frac{m_{i}}{\pi M_{\phi }^{2}}%
y_{ei}^{-}y_{\mu i}^{-}I(r)  \label{A-a}
\end{eqnarray}%
where $r\equiv \frac{m_{i}^{2}}{M_{\phi }^{2}}-1$ and the two functions
being:%
\begin{eqnarray}
G(r) &=&\frac{1}{3r}+\frac{3}{2r^{2}}+\frac{1}{r^{3}}-\frac{\left(
1+r\right) ^{2}}{r^{4}}\ln (1+r)  \label{G(r)} \\
&&  \nonumber \\
I(r) &=&\frac{1}{r}+\frac{2}{r^{2}}-\frac{2}{r^{2}}\ln (1+r)-\frac{2}{r^{3}}%
\ln (1+r)
\end{eqnarray}

In Fig 2(b) we have intermediate states of a neutral scalar boson and a
charged lepton. The amplitudes are%
\begin{eqnarray}
A_{+}^{\phi }\left( b\right) &=&\sum_{i}\frac{m_{\mu }}{\pi M_{\phi }^{2}}%
y_{ei}^{+}y_{\mu i}^{-}H\left( r\right) +\sum_{i}\frac{m_{i}}{\pi M_{\phi
}^{2}}y_{ei}^{+}y_{\mu i}^{+}K\left( r\right)  \label{A+b} \\
&&  \nonumber \\
A_{-}^{\phi }\left( b\right) &=&\sum_{i}\frac{m_{\mu }}{\pi M_{\phi }^{2}}%
y_{ei}^{-}y_{\mu i}^{+}H\left( r\right) +\sum_{i}\frac{m_{i}}{\pi M_{\phi
}^{2}}y_{ei}^{-}y_{\mu i}^{-}K\left( r\right)  \label{A-b}
\end{eqnarray}%
where the two functions are%
\begin{eqnarray}
H(r) &=&\frac{1}{6r}-\frac{1}{2r^{2}}-\frac{1}{r^{3}}+\frac{1+r}{r^{4}}\ln
\left( 1+r\right) \\
&&  \nonumber \\
K(r) &=&\frac{1}{r}-\frac{2}{r^{2}}+\frac{2}{r^{3}}\ln (1+r)
\end{eqnarray}

We shall from now ignore the $A_{-}$ amplitudes as they are similar to the $%
A_{+}$ results. After making the simplifying assumption that the masses of
the charged scalar boson $M_{\phi }$ and neutral lepton $m_{i}$ of Eq (\ref%
{A+a}) are the same as those for the neutral scalar boson and charged lepton
of Eq (\ref{A+b}), we add the two amplitudes from (\ref{A+a}) and (\ref{A+b}%
):\ $A_{+}^{\phi }\left( a\right) +A_{+}^{\phi }\left( b\right) =$ $A_{+}.$
Various mass limits as shown in Eqs (\ref{limit-a}) - (\ref{limit-c}) can be
taken in a straightforward manner and we display the results\footnote{%
Subleading terms are dropped from the results in Table 2. Also, in the $%
m_{i}\gg M_{\phi }$ amplitude, the leading $1/m_{i}$ terms from Figs 2(a)
and 2(b) actually cancel if the Yukawa couplings for the charged and neutral
scalar bosons are identical. Since there is no reason to expect such an
equality, we keep one of these dominant terms.} in Table 2. We have also
listed, in the third column, the results when we sum over the contribution
of the whole tower of KK states\cite{towerSUM} according to the simple
one-extra-dimension formula of $M\left( n\right) =n/R=nM\left( 1\right) $
when $M\left( 0\right) \simeq 0.$ Our purpose is to demonstrate that no
qualitatively new feature appears in such amplitude sums, which give overall
numerical coefficients and retain the same mass dependences.%
\begin{eqnarray*}
&&\text{ \ \ \ \ \ \ \ \ \ \ \ \ \ \ \ \ \ \ \ Table 2.\ \ The scalar loop
amplitudes~\ }A_{+} \\
&&%
\begin{tabular}{|c|c|c|}
\hline
limits & $A_{+}$ & $A_{+}^{\left( sum\right) }$ \\ \hline
$m_{i}\gg M_{\phi }$ & $\ \ \ \ \ \ \ \ \ \ \sum_{i}\frac{1}{\pi m_{i}}%
y_{ei}^{+}y_{\mu i}^{+}\ \ \ \ \ \ \ \ \ \ $ & $36.8\sum_{i}\frac{1}{\pi
m_{i}\left( 1\right) }y_{ei}^{+}y_{\mu i}^{+}$ \\ 
&  &  \\ 
$m_{i}\ll M_{\phi }$ & $\ \ \ 4\sum_{i}\ln \left( \frac{M_{\phi }}{m_{i}}%
\right) \frac{m_{i}}{\pi M_{\phi }^{2}{}}y_{ei}^{+}y_{\mu i}^{+}\ $ & $\ \ \ 
\frac{2\pi }{3}\sum_{i}\ln \left[ \frac{M_{\phi }^{{}}\left( 1\right) }{m_{i}%
}\right] \frac{m_{i}}{M_{\phi }^{2}\left( 1\right) }y_{ei}^{+}y_{\mu i}^{+}$
\ \  \\ 
&  &  \\ 
$m_{i}\approx M_{\phi }$ & $\ \ \ \ \ \frac{1}{3}\sum_{i}\frac{1}{\pi
M_{\phi }{}}y_{ei}^{+}y_{\mu i}^{+}\ \ \ \ \ \ $ & $12.47\sum_{i}\frac{1}{%
\pi M_{\phi }\left( 1\right) }y_{ei}^{+}y_{\mu i}^{+}$ \\ \hline
\end{tabular}%
\end{eqnarray*}

The amplitude $A_{+}$ in the $m_{i}\simeq M_{\phi }$ case, as well in the $%
m_{i}\gg M_{\phi }$ limit, has only one power of heavy mass in the
denominator --- they are as large as allowed by the decoupling theorem,
which requires the amplitude to vanish when the heavy mass approaches
infinity. The $m_{i}\ll M_{\phi }$ amplitude is somewhat more damped, by the
heavy scalar mass $M_{\phi }$ as $m_{i}M_{\phi }^{-2}\ln M_{\phi }$.

\section{Discussion}

\subsection{The chiral symmetry perspective}

The structure of the $\mu e\gamma $ amplitude in Eq (\ref{amplitude}),
symbolically written as $\bar{\psi}_{L}\sigma \psi _{R}F$ or $\bar{\psi}%
_{R}\sigma \psi _{L}F,$ involves flipping the fermion chirality. Thus the
amplitude must be proportional to a fermion mass. Before discussing details,
we observe that if we have bulk leptons propagating in the extra dimensions,
then chiral symmetry is broken in the effective four dimensional theory by
their Kaluza-Klein states, which are necessarily massive. Having a large
chiral symmetry breaking, such theories offer from the outset the
possibility for a less suppressed $\mu e\gamma $ amplitude. This statement
is valid whether the higher dimensional theory has chiral symmetry or not.
Our basic assumption is that the zero mode fermion masses are negligibly
small compared to their KK excitation. The largest possible fermion mass
that can bring about the chirality change in the $\mu e\gamma $ amplitude is
different in the vector and scalar loop diagrams.

For the vector loop contribution, we have assumed that there are only
left-handed charged current couplings\footnote{%
We have not considered neutral vector loop case as such diagrams would
involve further suppressions at the flavor-changing vertices.}. In such a
situation, helicity change takes place on the external lepton lines ---
hence the relevant mass is that of the muon. Since the amplitude corresponds
to a dimension-five operator, it must have an overall dimension of
inverse-mass. Thus, in the vector loop amplitude, we expect a damping factor
of $m_{\mu }/M_{W}^{2}$ as shown in Eq (\ref{W-amplitude}). If this was the
principal suppression factor, the resultant amplitude and decay rate would
still be large. The unitarity condition for the mixing matrices $%
\sum_{i=1}^{3}\mathbb{U}_{ei}^{\ast }\mathbb{U}_{\mu i}=0$ causes the actual
amplitude to be much more damped as the subleading term of the $F$-function
in Eq (\ref{F-z}) generally has two additional powers of heavy masses in the
denominator. Namely, a form of GIM mechanism\cite{GIM} is operative here.

The scalar loop amplitudes are less suppressed for two reasons: (1) here the
necessary chirality change can be effected by the large intermediate lepton
mass, and (2) in the scalar case there is no cancellation mechanism $%
\sum_{i=1}^{3}\mathbb{U}_{ei}^{\ast }\mathbb{U}_{\mu i}=0,$ as in the vector
category, to further suppress the amplitude. For the $m_{i}\gg M_{\phi }$
case, the heavy lepton mass $m_{i}$ in the numerator flips the helicity, and
its propagator provides two powers of $m_{i}$ in the denominator, giving an
over all $1/m_{i}$ suppression. For the $m_{i}\approx M_{\phi }$ case,
either the scalar or lepton propagator can provide the mass power in the
denominator. Because their masses are comparable, the resultant suppression
is again $1/m_{i}.$

\subsection{A numerical estimate of the $\protect\mu e\protect\gamma $
branching ratio}

We find it particularly interesting that the $\mu e\gamma $ amplitude can be
less suppressed when the intermediate scalar and lepton masses are
comparable, leading to a possibly observable decay rate. From our experience
with the SM, we expect the Yukawa coupling to be small, on the order of
gauge coupling times the (zero mode) mass ratio of lepton over gauge boson.
In particular, there has been the suggestion of neutral scalar's coupling to
two charged fermions ($i$ and $j$) being on the order of $g\sqrt{m_{i}m_{j}}%
/M_{W}.$ This coupling ansatz\cite{ChengSher} has been studied extensively,
and found to be compatible with known phenomenology. With this estimate of
the Yukawa strength, the loop in Fig 2(b) with the intermediate states being
the KK states of a tau-lepton and a scalar boson would yield a branching
ratio of%
\begin{eqnarray}
B\left( \mu e\gamma \right) &\simeq &\frac{\alpha }{\pi g^{4}}\left( \frac{%
M_{W}^{4}}{m_{\mu }^{2}M_{\phi }^{2}}\right) \left( y_{e\tau }^{+}y_{\tau
\mu }^{+}\right) ^{2}  \nonumber \\
&\simeq &\frac{\alpha }{\pi }\left( \frac{m_{e}}{m_{\mu }}\right) \left( 
\frac{m_{\tau }}{M_{\phi }}\right) ^{2}.  \label{estimate}
\end{eqnarray}%
For the first excited KK state with $M_{\phi }=O\left( TeV\right) $ and $%
m_{e,\mu ,\tau }$ being zero mode lepton masses, Eq (\ref{estimate}) gives a 
$B\left( \mu e\gamma \right) =O\left( 10^{-11}\right) ,$ which is comparable
to the current experimental limit \cite{muegammalimit} of $\ B\left( \mu
e\gamma \right) \lesssim 1.2\times 10^{-11}.$ This means that it's entirely
conceivable that the rate predicted by the scalar loop effect is within
reach of experimental detection in the near future\cite{mu-e-g-proposal}.

\subsection{Tau decays}

The considerations in this paper can be applied directly to radiative decays
of the tau lepton: $\tau \rightarrow \mu \gamma $ and $\tau \rightarrow
e\gamma .$ To the extent that the final lepton mass being negligibly small
compared to the initial lepton mass, the tau decay results involve replacing
the muon mass $m_{\mu }$ in equations such as Eq (\ref{W-amplitude}) by the
tau lepton mass $m_{\tau }.$ Thus an estimate of the branching ratio for the 
$\tau \mu \gamma $ decay yields%
\begin{eqnarray}
B\left( \tau \mu \gamma \right) &\simeq &0.17\times \frac{\alpha }{\pi g^{4}}%
\left( \frac{M_{W}^{4}}{m_{\tau }^{2}M_{\phi }^{2}}\right) \left( y_{\mu
\tau }^{+}y_{\tau \tau }^{+}\right) ^{2}  \nonumber \\
&\simeq &0.17\times \frac{\alpha }{\pi }\left( \frac{m_{\mu }m_{\tau }}{%
M_{\phi }^{2}}\right) \\
&=&O\left( 10^{-10}\right) .
\end{eqnarray}%
where $0.17$ is the branching ratio $B\left( \tau \rightarrow \mu \nu \bar{%
\nu}\right) $ and we have used the estimates of $y_{\tau \tau }^{+}=gm_{\tau
}/M_{W}$ and $y_{\mu \tau }^{+}=g\sqrt{m_{\mu }m_{\tau }}/M_{W}.$ Since the
existing limit\cite{tau-mu} $B\left( \tau \mu \gamma \right) =1.1\times
10^{-6}$ is still much larger than this estimate, it suggests that such a
discovery would only come about after a significant increase in experimental
detection efficiency has been achieved. Since our model estimate suggests $%
B\left( \tau e\gamma \right) \ll B\left( \tau \mu \gamma \right) ,$ it is
even less likely that the decay\cite{tau-e} $\tau \rightarrow e\gamma $
would be uncovered any time soon.

\subsection{Conclusion}

We have investigated the dependence by the $\mu e\gamma $ amplitude on heavy
particle masses, finding marked difference between vector and scalar loop
contributions. The vector loop amplitude, even in the comparable mass case,
is strongly suppressed by powers of neutrino mass divided by the heavy mass,
leading to such a small rate that the decay is predicted to be unobservable
in the foreseeable future. We have calculated the precise mass-dependence of
the scalar loops. The loop amplitude with a single heavy fermion is less
suppressed, not surprisingly, with one power of heavy fermion mass in the
denominator. Interestingly, scalar loop amplitudes with approximately equal
intermediate scalar and fermion masses (as, for example, the case in the
universal extra dimension theory) are also less suppressed. Calculation
using a plausible model of Yukawa couplings shows that such linearly damped
amplitudes can lead to a decay rate accessible by the next generation of $%
\mu e\gamma $ experiments.

B.H., T.P.C. and L.F.L. acknowledge the respective support from the
University of Missouri Research Board and from U.S. Department of Energy
(Grant No. DE-FG 02-91 ER 40682). T.P.C. also would like to thank Julia
Thompson, Simon Eidelman, David Kraus, and Peter Truoel for information on
the current experimental proposals for the relevant decay processed
discussed in this paper.

\bigskip 

\begin{center}
\underline{{\Large Figure captions}}
\end{center}

\begin{quotation}
\textbf{Fig 1.\ \ \ }$\mu \rightarrow e\gamma $ as mediated by a vector
loop. Contributions by diagrams with photon emitted by the external leptons
must also be included in the calculation.

\textbf{Fig 2.}\ \ \ \textbf{\ }$\mu \rightarrow e\gamma $ as mediated by a
loop having (a) a charged scalar, and (b) a neutral scalar, boson.
\end{quotation}

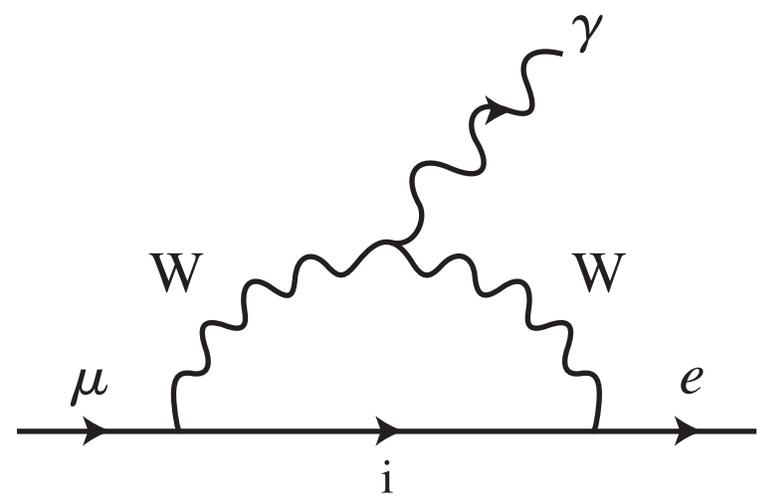
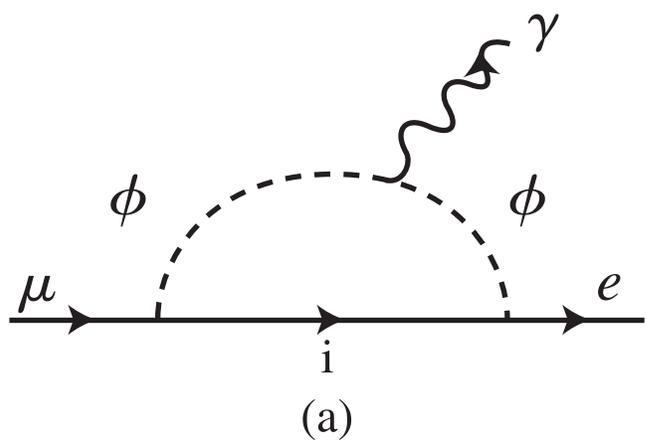
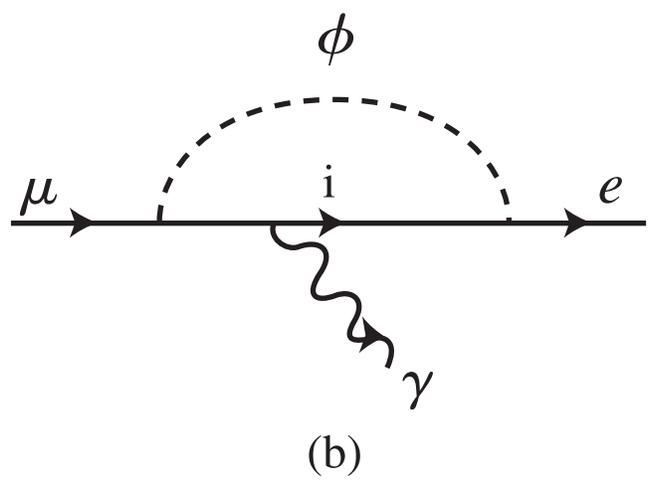
\end{document}